# Energy Consumption of Plant Factory with Artificial Light: Challenges and Opportunities

Wenyi Cai[1,2,†], Kunlang Bu[1,2,†], Lingyan Zha[3], Jingjin Zhang[3], Dayi Lai[4], Hua Bao[1,2,*]

[1]Global Institute of Future Technology, Shanghai Jiao Tong University, Shanghai 200240, China
[2]University of Michigan-Shanghai Jiao Tong University Joint Institute, Shanghai Jiao Tong University, Shanghai 200240, China
[3]School of Agriculture and Biology, Shanghai Jiao Tong University, Shanghai 200240, China
[4]School of Design, Shanghai Jiao Tong University, Shanghai 200240, China

**ABSTRACT**

Plant factory with artificial light (PFAL) is a promising technology for relieving the food crisis, especially in urban areas or arid regions endowed with abundant resources. However, lighting and HVAC (heating, ventilation, and air conditioning) systems of PFAL have led to much greater energy consumption than open-field and greenhouse farming, limiting the application of PFAL to a wider extent. Recent researches pay much more attention to the optimization of energy consumption in order to develop and promote the PFAL technology with reduced energy usage. This work comprehensively summarizes the current energy-saving methods on lighting, HVAC systems, as well as their coupling methods for a more energy-efficient PFAL. Besides, we offer our perspectives on further energy-saving strategies and exploit the renewable energy resources for PFAL to respond to the urgent need for energy-efficient production.
**KEYWORDS:** Plant factory with artificial light, energy saving, lighting, HVAC, renewable energy

[†]These authors contributed equally to this work.
[*]Corresponding author. E-mail address: hua.bao@sjtu.edu.cn (H. Bao).

# Introduction

A series of pressures, including rapid population growth, urbanization, climate change, land degradation, and biodiversity loss related to extreme weather events, challenge the ability of food systems to provide sufficient and steady food supply. However, the open-field farming system cannot fully meet the requirements of the growing population due to the extremely long and costly supply chain and the nonuniformity of freshwater resources [1-4]. Figure 1 shows the global distribution of prediction of crops yield in 10 years per capita of both national and subnational survey regions. The prediction shows that South America and North America will have high yields, while most parts of Africa and the Middle East will still have food scarcity. In order to relieve the food crisis all over the world, a new form of agricultural cultivation has attracted much attention: An indoor plant factory system with artificial light (PFAL) for efficient production of crops (Fig. 2A).

The advantages of PFAL include high crop yields, efficient use of water and nutrients, reduced need for pesticides, and year-round continuous production [5-8]. Additionally, because PFAL is often located close to urban populations, it can reduce transportation costs and carbon emissions compared to open-field farming and greenhouse farming (Fig. 2B) [9, 10]. PFAL can be used to grow a large variety of crops, including leafy greens, herbs, fruits, and vegetables by providing suitable growing conditions. Besides, this technology has rapidly advanced in recent years with the progress of energy-efficient LED lighting [11]. However, the energy consumption and the resulting running cost of PFAL are still very high compared to open-field and greenhouse farming [11], which limits its global expansion.

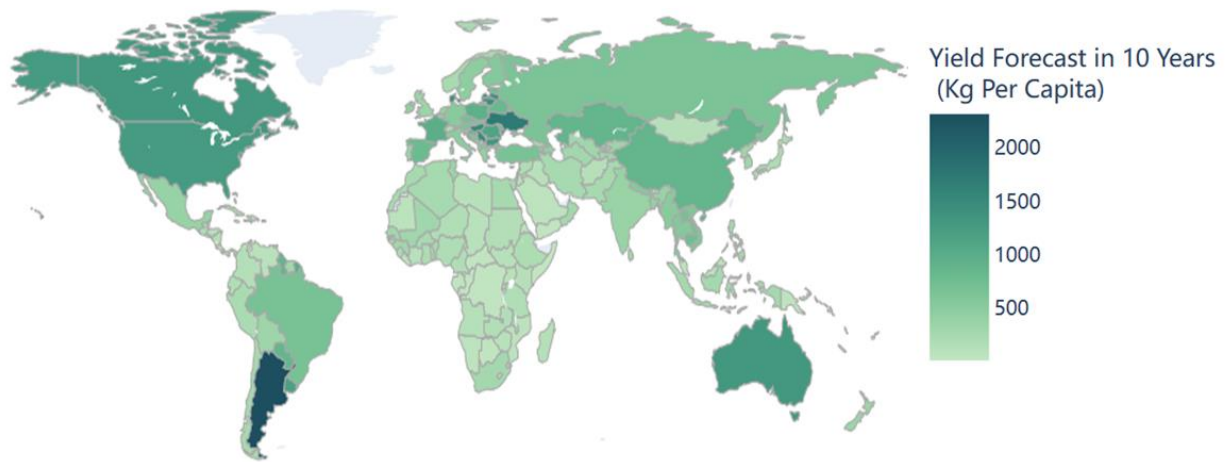

Figure 1. Geography of the global yield forecast in 10 years (kg crops per capita). Obtained by applying ARIMA model to annual yield and population data, from 1961 to 2022 from the Food and Agriculture Organization (FAO) and the Population Division Apartment of United Nations, respectively.

Some review papers and books have covered various aspects of the research and development of PFAL. For example, Kozai and coauthors responded to the increasing interest and demand for PFAL by providing a broad vision on environmental and resource issues, research and development history, cultivation of plants, as well as design and management of PFAL [12]. Delden et al. gave their perspective on the product quality, automation, robotics, system control and environmental sustainability of PFAL and how research and development, socio-economic and policy-related institutions work together to ensure a successful PFAL system [7]. In terms of lighting devices, Liu et al. reviewed recent progress aiming to improve light conversion efficiency and nutritive properties of crops by different lighting strategies, providing economic irradiation patterns or modes for various PFALs production requirements [13]. Engler et al. reviewed the economic feasibility and environmental impacts of various controlled environmental agricultural facilities (such as GHs and PFs) [14]. Recently, due to the global challenge of net zero carbon emission combined with the growing energy price, there has been growing interest in increasing the energy use efficiency (the amount of marketable product per unit of electricity) of PFAL. However, a review dedicated to the energy consumption for PFAL is still lacking.

In this review, we provide a comprehensive summarization on the recent advances in improving the energy use efficiency of PFAL and offer our perspectives on future research, especially the possibility of combining PFAL with renewable energy technology. This manuscript is organized as follows. We first provide a brief introduction to PFAL as well as its reported energy consumption. We then review the current energy-saving methods in PFAL, especially lighting and HVAC systems. Lastly, we propose our perspectives on the future trends for energy-saving methods from various engineering fields and how to better integrate the PFAL with renewable energy technologies and advanced energy materials.

**Energy Consumption of PFAL**

A conventional PFAL consists of six principal structural elements (Fig. 2C) [15]: (1) thermally insulated and opaque envelope with vertical farming units inside; (2) A lighting system such as fluorescent light or light-emitting diodes (LEDs) over the culture units to supply lighting sources; (3) A HVAC system including heating, ventilation and air conditioning systems used for controlling the indoor temperature and humidity around plants in the culture room, and circulating air to enhance photosynthesis and transpiration; (4) A $CO_2$ supply unit to maintain $CO_2$ concentration in the culture room around plants to ensure the plant photosynthesis; (5) A nutrient solution and water supply unit; (6) An environmental control unit including humidity detection, electric conductivity detection and pH controllers for the nutrient solution.

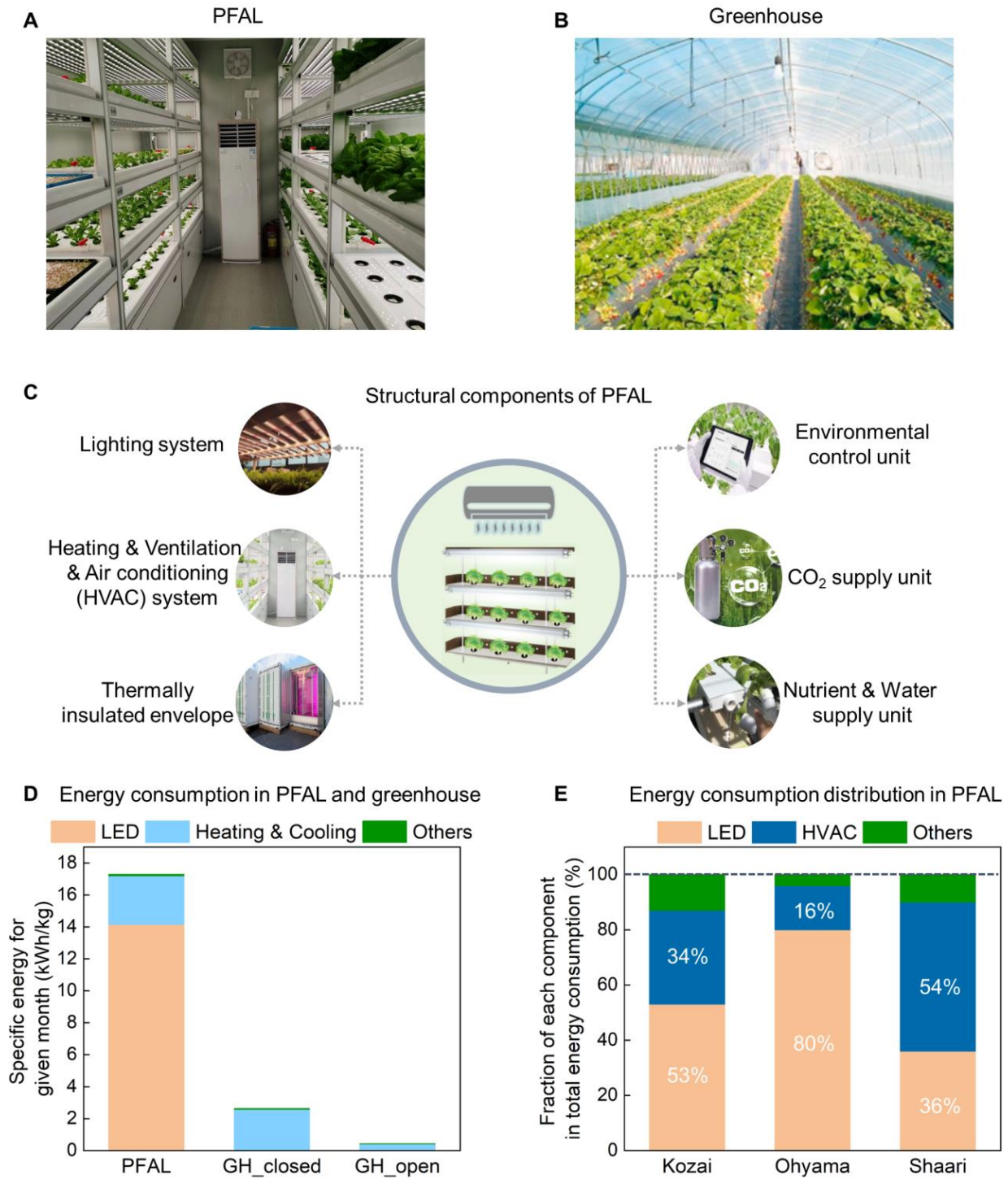

Figure 2. (A) Photograph of a typical PFAL. (B) Photograph of a typical greenhouse [9]. (C) Schematic view of the structure of plant factory, with six principal structural elements of PFAL including lighting devices, nutrient and water supply unit, $CO_2$ supply unit, thermally insulated

envelope, environment control unit, heating, ventilation, and air conditioning (HVAC) system [15]. (D) Specific energy consumption of three farming systems for a typical day in July (summer) [16]. The difference in energy consumption between greenhouse farming and PFAL shows the high energy consumption of PFAL. (E) Energy consumption distribution of PFAL for different researchers [17-19]. The different distribution proportions show the importance of lighting and HVAC systems in the total energy consumption of PFAL.

It is generally known that the energy consumption in a PFAL is much higher than both the open-field farming system and the greenhouse system [9, 20]. For example, Graamans et al. found that PFAL consumed more electricity than greenhouse [21]. Weidner et al. showed that the energy consumption of PFAL in Stockholm is much higher than that of open and enclosed greenhouses (Fig. 2D), especially in summer. Such high energy consumption in PFAL is mainly due to the LEDs and HVAC systems [16]. More importantly, although PFAL has a higher energy consumption than a greenhouse, the productivity of PFAL is also much higher. Therefore, regarding the energy use efficiency for crop production, there is still a debate on whether the PFAL is better than greenhouses. For example, Graamans et al. have found that PFAL is more energy efficient than greenhouse [21] while Harbick et al. found the energy use efficiency of PFAL is inferior to all greenhouse systems [22]. As a result, further optimization of energy use efficiency is critical to the future application of PFAL.

While the PFAL has high energy consumption and all these six structural elements consume electricity, the distribution varies greatly (Fig. 2E): Kozai and Yokoyama's group reported that the energy consumption of LED accounts for 53%, the energy consumption for HVAC accounts for 34% and the rest accounts for 13% in PFAL [17]. Ohyama et al. reported that lighting energy consumption accounts for 80%, the energy consumption for HVAC accounts for 16%, and the others account for 4% [18]. Shaari et al. reported that the HVAC energy consumption accounts for 54%, the lighting energy consumption accounts for 36%, and the others account for 10% [19]. The differences among these results can be due to the different local climates, envelope design, and the operation strategy. In all these researches, it is quite clear that the lighting and the HVAC system energy consumption occupy the largest fraction, which indicates these two factors need to be optimized to improve the energy use efficiency of PFAL.

# Energy Consumption Optimization Methods of PFAL

In the subsequent discussions, we mainly focus on the lighting system and the HVAC system which are two main energy consumption sources of PFAL. Note that reducing energy consumption can be achieved by two different strategies. One is to achieve a certain environmental condition with better engineering design and control. The other is to find more energy-saving environmental conditions for plants. In fact, substantial improvements in energy consumption optimization of the lighting system and the HVAC system have been attempted. In addition to a large improvement in energy use efficiency, this advancement can also be demonstrated by a large increase in the yield or weight of production with the same energy input.

**Lighting**

*Equipment level*

*LED light utilization* Lighting system is the most important in PFAL, where the light source input is totally supplied by the lighting equipment called artificial light. As a result, improvements on lighting equipment are important for lighting energy consumption optimization. Fluorescent lamp was used for illuminating the cultivated plants in PFAL in early times with a relatively low conversion efficiency (around 0.25) from electrical energy to photosynthetically active radiation (PAR, 400-700 nm) energy [15]. With the development of the LED technology, the conversion efficiency from electrical energy to PAR energy has been increased to 0.3-0.4, because the lighting spectrum of LED can be controlled in the PAR range (Fig. 3A) [23]. Accompanied by the increasing conversion efficiency, the photosynthetic photon efficacy (defined as the output number of photons to the input electrical energy) now become much higher, and can thus greatly improve the energy use efficiency [24, 25]. Recent measurements of the efficacy of horticulture lighting show that PAR efficacies mostly range from 1.3 to 2.1 $\mu mol \cdot J^{-1}$ through advances in LED lighting technology [26, 27], while high efficacy up to 4.0 $\mu mol \cdot J^{-1}$ can be achieved with some manufacturers. Besides, the emitting light distribution from a fluorescent lamp is quite uniform, and the light always illuminates the gaps between the nearby plants. In recent years, further progress has been achieved on LED technology, which can better control the light emission direction (Fig. 3B) [28, 29]. Precise illumination, which means to illuminate the cultivated plants below with a specific emission range and direction for totally illuminating the photons to the leaf areas, can efficiently improve the light energy use efficiency. The height and illumination range of LEDs can also be dynamically tuned with the growth of the plant [30, 31]. Besides, the outermost leaf area always lacks light irradiation because of the lighting emission direction. Arcel et al. set up an upward lighting system in real time to meet the optimal growth conditions of the crop (Fig. 3C) so that the supplemental lighting retarded the senescence of outer leaves and decreased waste [28, 29]. As a result, it led to an improvement on light energy use efficiency for acquiring higher marketable leaf fresh weight with the same energy input. It should also be mentioned that this precise illumination may require additional equipment and may lead to additional cost.

*Light intensity and spectrum control* The intensity of the lighting source is an important issue in helping cultivate the underneath plants. Fu et al. proved that the light energy use efficiency around 200 μmol·m$^{-2}$·s$^{-1}$ treatment was the highest for Romaine lettuce [32]. Besides, the light intensity also has to be associated with the growth stage of the plant. Zheng et al. suggested that the light intensity of 90 μmol·m$^{-2}$·s$^{-1}$ at the rooting stage and 270 μmol·m$^{-2}$·s$^{-1}$ at the seedling stage (Fig. 3D) [33]. As a result, the root dry weight of the strawberry can be increased by 106.1%. Except for the intensity of LED, the emission spectrum can also be precisely controlled. For example, Ji et al. have found that far-red radiation can help increase the dry mass of tomatoes by 26–45% [34]. Johkan et al. have proved that the blue light irradiation of seedlings can improve the seedling quality and growth after transplanting in red leaf lettuce (Fig. 3E) [33, 35].

*Lighting mode selection* Lighting modes are also important when combined with the planting strategy [13]. Continuous lighting is a conventional lighting mode which can improve quality traits such as soluble sugars and Vitamin C [36, 37] and inhibit nitrate [38] by extending the duration of light. To achieve more efficient production, an alternate lighting mode is introduced as alternately irradiates plants with different light spectrums or spectral compositions. Shimokawa et al. have found more than 60% of the biomass of lettuce cultivars was achieved with alternate irradiation by red and blue light without extra energy consumption [39]. Moreover, the photoperiod can be controlled with LED, resulting in intermittent lighting strategies. For example, Avgoustaki et al. innovatively separated photoperiod into 14h/10h light/dark cycle [40]. As a result, the biomass of Ocimum basilicum was reported to increase by 47%, and the energy consumption was simultaneously decreased by 15.9% corresponding to the fluctuation of electricity prices. The photoperiod can also be changed corresponding to the ambient environment to reduce cooling/heating load [41, 42].

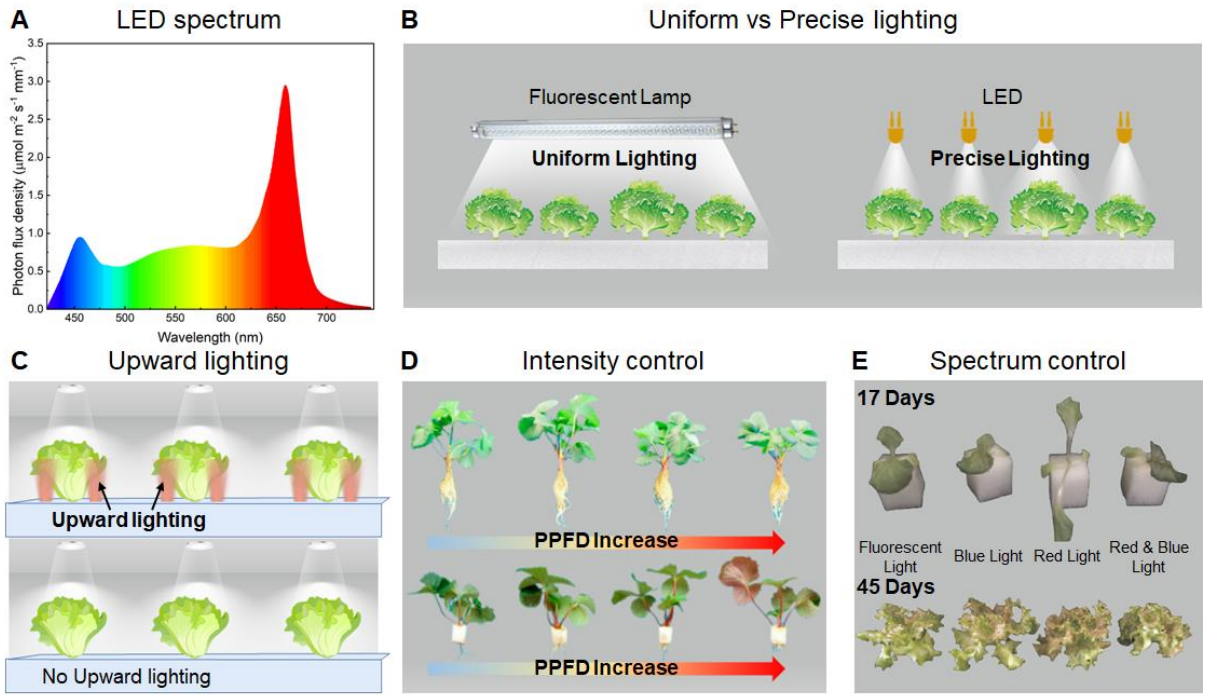

Figure 3. Energy saving of lighting system in PFAL.

(A) Spectral distributions of white LEDs with the red-blue ratio of 0.9 and 2.7 [23]. (B) Precise lighting function of LED instead of fluorescent lamp [28]. (C) Upward lighting results in an improvement of the marketable leaf fresh weight [28]. (D) Intensity control of LED on rooting and growth of hydroponic strawberry runner plants [33, 34]. (E) Spectrum control of LED to improve seedling quality and growth after transplanting [35].

*Plant level*

*Plant type selection* Although the PFAL mainly focus on the improvements in equipment level design in PFAL with the development of technology, improvements can be further realized by optimizing the light energy consumption on the plant level. First, it is important to select the proper plant type that meets the features of PFAL on both the lighting equipment and the controlled environment. A compact plant is more desirable for lighting energy saving with more photons emitted to the plant surfaces for absorption. Similarly, an open canopy with long internodes and narrow leaves is beneficial for uniform light distribution [43]. Besides, plants that can withstand harsh growth conditions and have high marketable prices are preferred so as to have high prices with the same lighting energy consumption.

*Planting strategy selection* Planting strategy is another option for lighting energy saving such as intercropping, which means simultaneous growing of multiple crops. For example, Li et al. have proved that yield benefits of intercropping increased through time [44]. Besides, the grain yields in intercropped systems were on average 22% greater than in matched monocultures and had greater year-to-year stability [45]. Although the intercropping technology is underexplored in plant factories so far, this strategy may be applied to PFAL to better enhance the light trapping capability of the plants, thus reducing lighting energy [7].

## HVAC

### Factory level

*Location selection* For HVAC system, energy consumption is highly related to the local climate [16, 20, 22, 42, 46, 47], and therefore choosing the proper location to build plant factories can save energy. Weidner et al. calculated the energy consumption of the same plant factory (10000 m$^3$ volume) in different locations [16]. They calculated the specific energy (kWh kg$^{-1}$) by dividing the simulated energy consumption (kWh) by the model prediction of lettuce production (kg). As the location moved from cold regions (e.g., Reykjavik region and Stockholm region) to hot regions (e.g., United Arab Emirates (UAE) and Singapore), the cooling demand increased sharply. For example, the specific energy for mechanical cooling in UAE (23.91° N) was more than 15 times that in Reykjavik region (64.15° N) (Fig. 4A). This resulted in the higher overall specific energy in UAE (20.1 kWh kg$^{-1}$) compared to the Reykjavik region (15.6 kWh kg$^{-1}$). Moreover, Song et al. [42] found a ~7.7 kWh daily electricity consumption difference for the same 20-foot container PFAL in Haikou (~20° N, very hot zone) and Mohe (~50° N, subarctic and arctic zone) in China. Other research also discovered energy consumption differences for the same plant factory in the United States, Sweden, Netherland, and United Arab Emirates [20, 22, 46, 47]. When the located city cannot be changed and cooling accounts for the major part of energy consumption, utilizing existing shading such as trees or buildings, can also reduce cooling demand.

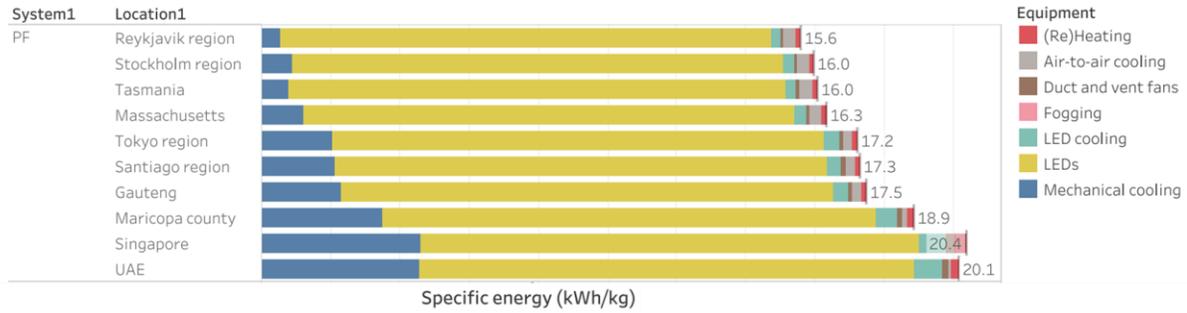

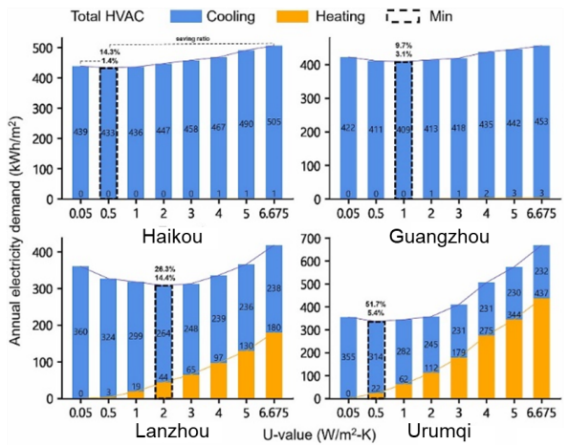

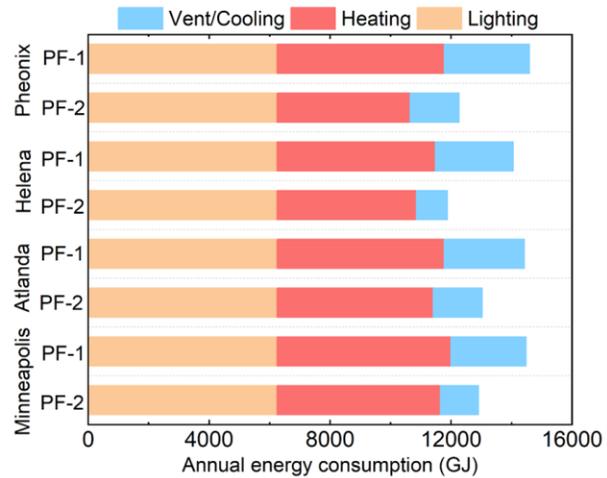

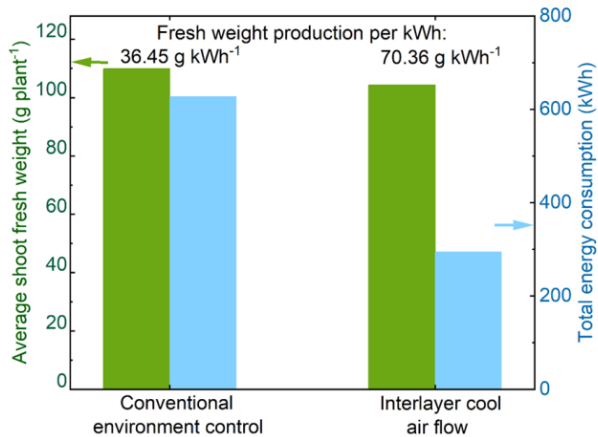

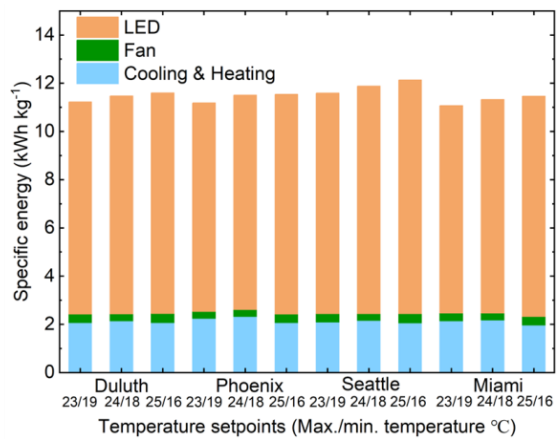

Figure 4. Energy saving of HVAC for PFAL.

(A) Specific energy consumption (kWh kg$^{-1}$) of plant factories in different locations [16]. (B) Annual electricity consumption for a 20-foot container plant factory with different overall heat transfer coefficients (U) at four locations (a total of eight locations were calculated in the original paper, only four locations are shown here for simplicity) [42]. (C) Energy saving of a plant factory (29.3 m × 58.5 m × 3.4 m) with (PF-2) and without (PF-1) using economizers [22]. (D) Energy saving per fresh weight for plant factories with interlayer cool airflow ventilation compared to conventional environment control ventilation [48]. (E) Energy use efficiency for the plant factory (6.03 m × 3.71 m × 3.91 m) choosing different temperature setpoints in different locations [46].

*Envelope design* For the plant factory, the envelope mainly indicates the four sidewalls (or façade), the roof, and the floor. Traditional plant factories tend to enhance the insulation of the envelope so that agriculture is less dependent on the environment [47, 49]. However, increasing insulation (or reducing the overall heat transfer coefficient) may increase energy consumption since heat generation inside the plant factory is difficult to be dissipated by the envelope, thus the cooling demand increases. Song et al. [42] calculated the electricity consumption for a 20-foot container plant factory with different overall heat transfer coefficients (U) from 0.05 ~ 6.675 W m$^{-2}$ K$^{-1}$ at different locations (Fig. 4B). When increasing the overall heat transfer coefficient, the HVAC energy consumption first decreased and then increased. Therefore, an optimum heat transfer coefficient exists for the minimum energy consumption of a plant factory at a specific location. An optimum energy consumption occurred at U = 0.5 W m$^{-2}$ K$^{-1}$ in Haikou (~20° N, very hot zone) and Urumqi (~ 45° N, cold zone), while it occurred at U = 2 W m$^{-2}$ K$^{-1}$ in Lanzhou (~35° N, cool zone). The HVAC energy saving in PFAL when reducing insulation ranged between 1.4%~14.4% for different locations. Note that the target heat transfer coefficient for low energy consumption depends on the local climate, heat load, and size of PFAL. For a different PFAL in different locations, the energy consumption may continuously decrease with a higher heat transfer coefficient of the envelope [47].

The envelope heat absorption of PFAL can also be reduced by increasing surface albedo. Graamans et al. calculated the energy consumption of the plant factory with a floor area of 1296 m$^2$ and 3.5 m height when increasing surface albedo for three regions: United Arab Emirates (UAE), Netherland, and Sweden [47]. Increasing surface albedo from 0.1 to 0.9 can decrease the energy consumption in warm regions, including Netherland (-0.5%) and United Arab Emirates (-7.5%) due to reduced cooling load while an energy increase was found in Sweden (+0.8%) because of additional heating demand when solar heating is reduced [47].

From the above analysis, we can conclude that for the envelope design of PFAL, specific optimization analysis should be conducted based on specific locations and PFAL settings. Enhancing heat dissipation to reduce cooling demand in hot seasons is usually at the expense of increasing heating demand in cold seasons. If heat dissipation is excessively enhanced through envelope design, heating demand will increase rapidly and overall energy consumption will increase.

*Equipment level*

*Adding air-side economizer* An air-side economizer is a component of HVAC systems that helps to improve energy use efficiency by using outdoor air to cool/heat and ventilate indoor spaces [50]. Harbick et al. calculated the energy consumption of the plant factory (29.3 m × 58.5 m × 3.4 m) with or without using economizers in Phoenix, Helena, Atlanta, and Minneapolis (Fig. 4C) [22]. Both heating and cooling energy have been reduced in four locations. For example, the annual heating and cooling energy for a plant factory in Phoenix have been reduced by ~180 kWh m$^{-2}$ and ~195 kWh m$^{-2}$ when adding air-side economizers. The total HVAC energy consumption has been reduced by 27.8%, 27.8%, 17.1%, and 19.0% for plant factories in Phoenix, Helena, Atlanta, and Minneapolis. Other simulation results also found that introducing air-side economizer reduced the HVAC energy by 40-69% and 3.5%-52% in five locations in the United States and eight locations in China [24, 42].

*Ventilation regulation system design* Ventilation equipment refers to fans, ducts and other equipment that promote indoor/outdoor air exchange and room air distribution. Current ventilation system includes intake and/or exhaust fans installed on the wall plus additional fans above the plants [41, 42]. However, current design lacks detailed ventilation regulation for the micro-climate of the plants. If we focus more on the airflow organization of the micro-environment where the plant is located instead of the entire plant factory, it is expected that the energy consumption can be reduced. Li et al. [48] conducted an experiment to compare the energy consumption of interlayer cool airflow (micro-climate ventilation regulation) and the traditional overall ventilation technique in two plant factories in Beijing. Their ventilation regulation system introduced air into the interlayer between the cultivation board and the nutrient solution surface. Then the air flowed upward into the internal canopy through vent holes in the cultivation board to achieve ventilation. The energy consumption and fresh weight of the plant are shown in Fig. 4D. The electricity for air conditioning using optimized ventilation regulation systems significantly reduced (294.4 kWh) compared with the traditional ventilation system (627.6 kWh), with similar lettuce production. The fresh weight production per kWh of electricity for air conditioning for optimized ventilation techniques (70.36 g/kWh) was nearly twice the production for traditional ventilation (36.45 g/kWh). Therefore, optimizing ventilation regulation for plant micro-environment has huge potential to reduce energy consumption.

*Plant level*

*Operating setpoints for plant growth* Considering the plant growth and local weather conditions, it is possible to select optimum operating setpoints (temperature and humidity) for low energy consumption [46, 51]. Zhang et al. calculated the energy consumption for a plant factory (6.03 m × 3.71 m × 3.91 m) with different temperature and humidity setpoints in Duluth, Phoenix, Seattle, and Miami [46]. When changing the temperature setpoint from 23/19 °C (a narrower range) to 25/16 °C (a broader range) for the plant factory, the HVAC energy consumption was reduced by 4.1%, 8.7%, 4.8%, and 9.3% in four locations due to lower cooling and heating demand. The energy consumption also decreased by 12.8% when changing the maximum humidity from 65% to 75% in Duluth since the dehumidification load was reduced. It is worth noting that changing the temperature and humidity setpoints may affect plant production. When applying lettuce growth model [52], the authors found that the annual lettuce production using a broader temperature range (25/16 °C) was reduced by 5.5% compared to a narrower one (23/19 °C). Consequently, the specific energy consumption (kWh kg$^{-1}$ lettuce fresh weight) increased by ~3% at a broader temperature range compared to the a narrower one (23/19 °C) in four locations (shown in Fig. 4E).

Although the specific energy for HVAC decreased from 2.54 to 2.43 kWh kg$^{-1}$ using a broader temperature setpoints in Phoenix, the energy for LED increased from 8.64 to 9.11 kWh kg$^{-1}$ since LED power remained the same when only changing temperature setpoints. Therefore, the total specific energy increased when using a broader temperature range (25/16 °C) compared to a narrower one (23/19 °C). Therefore, changing temperature setpoints in PFAL can achieve energy savings, but may at the expense of reduced yields. Moreover, potential energy saving can be obtained by changing the operating setpoints with different plant types, growth stages, locations, or seasons.

Table I. Summary of the energy saving methods in PFAL.

| Authors | Optimization method | Plant Category | Location | Energy saving proportion | | Electricity required for lettuce growth (kWh/kg) |
|---|---|---|---|---|---|---|
| | | | | Lighting | HVAC | |
| Weidner et al. [16] | Locating in cold and dry regions | Lettuce | Ten locations[c] | | 22.40% | 15.6-20.1 |
| Zhang et al. [20] | Locating in cold and dry regions | Lettuce | Six cities[a] | | 34%~47% | 8.3~14.2 |
| Harbick et al. [22] | Locating in cold and dry regions Adding air-side economizer | Lettuce | Four cities in the US[b] | | 6.50% 17.1%~27.8% | |
| Eaton et al. [24] | Improving LED efficacy Adding air-side economizer | Lettuce | New York City | 12%-42% | 40%-69% | 6.2-12 |
| Zheng et al. [33] | Intensity control with plant category | Strawberry | Beijing | 21.10% | | |
| Ji et al. [34] | Far-red radiation | Tomato | Wageningen | 26%-45% | | |
| Bian et al. [38] | Alternate lighting: red and blue light | Lettuce | Yamaguchi | 37.50% | | |
| Avgoustaki et al. [40]. | Intermittent lighting strategy: photoperiod settings | Ocimum basilicum | Athens | 15.90% | | |
| Song et al. [42] | Locating in cold and dry regions Increasing wall heat transfer coefficient Adding air-side economizer Switching photoperiod | Lettuce | Eight cities in China[d] | | 29.40% 1.4%~51.7% 3.5%~51.2% 6.5%~19.4% | |
| Graamans et al. [47] | Increasing wall heat transfer coefficient Increasing surface albedo | Lettuce | United Arab Emirates, Netherlands, and Sweden | | 12.1%~30.6% -0.8%~7.5% | 8.5~10 |
| Li et al. [48] | Ventilation regulation | Lettuce | Beijing | | 53.10% | 14.2 |

[a]Six cities are Duluth, Seattle, Phoenix, Miami, Abu Dhabi, Riyadh
[b]Four cities are Atlanta, Helena, Minneapolis, and Phoenix
[c]Ten locations are Reykjavik region, Stockholm region, Tokyo region, Massachusetts, Tasmania, Singapore, Santiago region, Maricopa county, Gauteng, and United Arab Emirates
[d]Eight cities are Haikou, Guangzhou, Shanghai, Beijing, Lanzhou, Urumqi, Harbin, and Mohe

## Conclusions

Plant factory with artificial light (PFAL) is a promising technology in agriculture due to high crop production, efficient use of water and nutrients, and year-round continuous production. However, it consumes much larger amount of energy than the traditional farming and the greenhouse, limiting the widespread of PFAL. In this review, we focused on the two major parts of energy consumption in PFAL: Lighting and HVAC systems and reviewed current energy-saving techniques in lighting and HVAC systems. The summary of energy-saving methods in PFAL is shown in Table I, in which the electricity required to produce the unit weight of plants is an important index to guide for future work on PFAL.

Energy-saving methods in lighting mainly focus on the optimization of light intensity, light spectrum and light period. Adjusting the light intensity to produce the same weight of plants can save the lighting system energy by ~20%. Modifying the LED light spectrum (including introducing far-red radiation and switching between red and blue light) can also improve yield and consequently, reduce energy consumption. Implementing different light period settings including intermittent lighting has also been proven to reduce lighting energy by more than 10%. Moreover, changing the lighting period according to the ambient environment is also beneficial to reduce cooling/heating load by ~20%.

The energy consumption reduction methods for HVAC systems exist in many aspects, including optimizing overall plant factory design, adding energy-saving equipment, and adjusting plant growth temperature and humidity setpoints. The plant factory located in cold and dry regions with optimized envelope design (overall heat transfer coefficient and surface albedo) tends to have lower operating energy consumption. The HVAC systems equipped with air-side economizers can reduce energy consumption by 3.5%~51.2% according to the local climate, and accurate local ventilation can also cut down the energy consumption by ~50% in PFAL. Increasing the upper and lower limits of temperature and humidity can also help reduce energy consumption, but it may be at the expense of reduced plant yield. These promising improvements will provide more crop yield with lower energy consumption to relieve the food crisis.

# Perspectives

With the rapid development in various engineering fields, new technologies can always be integrated with current PFAL to further improve the energy use efficiency or to use more renewable energy in PFAL.

*Lighting and HVAC systems*

For both systems, using more efficient equipment will be possible in the future. In lighting, the LED efficacy is estimated to reach ~4.7 µmol·$J^{-1}$ in 2050 [53], meaning that the power consumption of LEDs will be halved compared to the current LED light with ~2 µmol·$J^{-1}$ efficacy. The lifetime of LEDs will also be improved in the future. For HVAC systems, replacing the traditional motors for pumps or fans in HVAC systems with adjustable speed motors in appropriate applications may lead to 30-50% energy consumption reduction [54]. For the lighting system, more attention needs to be paid to study the interactions between plants and light to develop LEDs that are more suitable for specific plants. Focusing on plant response to light with different light intensities, light spectrums, and light periods will help us design more energy-saving LEDs and lighting strategies. In HVAC systems, the energy consumption can be potentially reduced by optimized control techniques, accurate airflow organization, and advanced dehumidification methods. Advanced control algorithms including model predictive control (MPC) and fuzzy control can be employed to reduce energy consumption in PFAL [55, 56]. Moreover, ventilation of the entire plant factory space may increase energy consumption. Although previous experiment has been focused on the energy reduction of ventilation regulation systems [48], more theoretical research is needed to determine the energy reduction potential brought about by accurate air distribution in PFAL. Moreover, since the dehumidification load is usually large in PFAL due to evapotranspiration, advanced dehumidification methods used in buildings including desiccant-based dehumidification and heat recovery ventilators can be employed in PFAL [57, 58].

*Renewable Energy*

Renewable energy cannot directly reduce the energy consumption of PFAL, but it can allow PFAL to use more "green energy" and potentially with less cost, which reduces carbon emission. Photovoltaic panels, wind turbines, or small nuclear power plants can be adopted as energy sources for plant factories [59-61] for off-grid operation. Or they can also be connected to the electricity grid (on-grid) to offset part of running costs in PFAL. On the other hand, plants do not have to grow in a continuous, stable lighting environment. Therefore, it is theoretically possible to use PFAL as an energy storage system to offset the fluctuation of renewable energy by tuning the lighting with the fluctuation of energy source. To date, there is little discussion on how to better combine renewable energy systems with PFAL. This is an area that is worthy of further exploration.

*Advanced Energy Materials*

PFAL can be better integrated with advanced energy materials that have been developed recently. Just as examples, phase change materials can be utilized in PFAL to store heat in the daytime and release heat at night. Adding phase change materials can potentially reduce cooling and heating demand at the same time, and they have been applied to maintain appropriate inner air temperature in greenhouses [62]. Radiative cooling materials achieve cooling by strongly reflecting sunlight and emitting electromagnetic waves in the atmospheric window (8-13 µm) to exchange heat with the cold universe [63-66], which can be utilized to reduce cooling demand in PFAL. Since radiative cooling materials have strong solar reflectivity, they can also function as a reflector, which can be installed around the cultivation rack to reflect light and increase the light interception and photosynthesis of crops [67].

## Author Information

Corresponding Author

*Email: hua.bao@sjtu.edu.cn. Tel: +86-21-34206765-5221

## Acknowledgement

This research is supported by the Science and Technology Commission of Shanghai Municipality [grant No. 23N21900300]. Also, we would like to extend our sincere appreciation to Danfeng Huang from the School of Agriculture and Biology at Shanghai Jiao Tong University for her expert opinions, which significantly enriched the content and quality of this review paper. Her insights and expertise have been instrumental in shaping the direction of our work.